\documentclass[
superscriptaddress,
a4paper,
onecolumn,
amsmath,amssymb,
aps,
prfluids,
floatfix,
longbibliography,
12pt,
tightenlines
]{revtex4-2}

\usepackage{graphicx}
\usepackage{dcolumn}
\usepackage{bm}
\usepackage[amssymb]{SIunits}

\usepackage[dvipsnames]{xcolor}
\usepackage{soul}

\begin{document}

\title{Ice melting in an oscillatory flow}

\author{Sof\'ia Angriman}
\email{sofia.angriman@univ-cotedazur.fr}
\thanks{Current affiliation: Université Côte d'Azur, CNRS, INPHYNI, 06200, Nice, France.}
\affiliation{Physics of Fluids Department, Max Planck Center for Complex Fluid Dynamics, and J. M. Burgers Centre for Fluid Dynamics, University of Twente, P.O. Box 217, 7500AE Enschede, The Netherlands}

\author{Tobias de Buck}
\affiliation{Physics of Fluids Department, Max Planck Center for Complex Fluid Dynamics, and J. M. Burgers Centre for Fluid Dynamics, University of Twente, P.O. Box 217, 7500AE Enschede, The Netherlands}

\author{Roberto Verzicco}
\affiliation{Gran Sasso Science Institute - Viale F. Crispi, 7 67100 L'Aquila, Italy}
\affiliation{Physics of Fluids Department, Max Planck Center for Complex Fluid Dynamics, and J. M. Burgers Centre for Fluid Dynamics, University of Twente, P.O. Box 217, 7500AE Enschede, The Netherlands}

\author{Sander G. Huisman}
\email{s.g.huisman@utwente.nl}
\affiliation{Physics of Fluids Department, Max Planck Center for Complex Fluid Dynamics, and J. M. Burgers Centre for Fluid Dynamics, University of Twente, P.O. Box 217, 7500AE Enschede, The Netherlands}

\begin{abstract}
We investigate the melting dynamics of an ice disk subjected to an external oscillatory flow using two-dimensional direct numerical simulations, in the absence of buoyancy, varying the flow amplitude and its oscillation frequency. We identify two distinct regimes governed by the interplay between advection and diffusion within the boundary layer. For slow oscillations, the melting process is well described by an effective steady flow, where a description based on classical forced convection is applicable. For fast oscillations the melting time increases significantly and approaches the diffusion limit regime, as a result of the oscillatory flow being unable to renew the fluid within the oscillating boundary layer, causing cold meltwater to accumulate near the interface and reducing heat transfer.

\end{abstract}

\maketitle

\section{Introduction}
Pulsating and oscillatory flows are common to many systems, including natural and engineering settings. 
In biological contexts, the blood flow in our arteries is dictated by the pumping of the heart \citep{Ku1997}, and similar oscillatory dynamics influence the circulation of cerebrospinal fluid \citep{Kelley2023}. The oscillatory nature of breathing introduces non-trivial transport phenomena during the transition between the successive respiratory cycles \citep{Lieber1998}. At the microscale, the cilia present in kidney cells respond to the pulsating nephrotic flow, affecting the generation of calcium \citep{Rydholm2010}.
In industrial applications, oscillatory flow reactors have attracted considerable interest due to their enhanced mixing capabilities, with applications in chemical and the pharmaceutical processing \citep{Mcglone2015}. 

In many of these systems oscillatory flows are intrinsically coupled to heat and mass transfer \citep{Cheng1998}. Examples include Stirling engines, internal combustion engines, and cryogenic devices, where heat exchange occurs in periodically reversing flows \citep{Choudhari2022}. At a more fundamental level, studies of oscillating bodies in fluid flows have shown that heat transfer can be modified by unsteady motion. For instance, the heat transfer from a heated cylinder oscillating in a cross flow can be enhanced when the oscillation frequency matches with the natural vortex shedding frequency \citep{Fu2002}, indicating a non-trivial coupling between heat transfer modification and oscillatory motion.

Oscillatory flows also play an important role in geophysical environments. Free-surface waves enhance heat transfer at the seabed boundary layer, with the frequency for optimal enhancement depending on ocean depth \citep{Michele2021}. Waves can also affect ice cliff erosion, and their effect on melt-rate decreases with depth \citep{Wolterman2026}. Similarly, the melting of ice in natural environments is strongly influenced by ambient flows \citep{Du2024}. While several studies have focused on steady currents \citep{Hester2021,Yang2024_JFM}, observations indicate that non-steady, such as tidal motions, are also relevant. Indeed, the short time scale variability of ice bodies melting in nature is due to the effect of ocean currents, and melting rates have been observed to be modulated by tidal currents \citep{Padman2018}. However, predictions are based on parameterizations that depend strongly on assumptions made on entrainment coefficients and ice geometry \citep{Anselin2023}. Controlled laboratory experiments of fully-submerged ice blocks vertically oscillating in saline water indicate that the oscillation strongly enhances the side melting rates, and can give rise to small-scale scalloping of the ice surface \citep{Sweetman2025}.

Despite these advances, a clear physical description of how oscillatory flows influence ice melting remains incomplete. In particular, it is not well understood under what conditions an oscillatory external flow enhances heat transfer, or conversely, when it becomes inefficient and heat transport becomes dominated by other mechanisms.
In this work, we address this question by studying the melting of an ice disk subjected to an external flow that oscillates periodically in time about a zero-mean by means of two-dimensional direct numerical simulations. 
This idealised configuration allows us to isolate the effect of unsteady forced convection. Throughout this study, buoyancy effects are neglected so that the analysis focuses on forced-convection dynamics.
By systematically varying the amplitude and frequency of the oscillating flow, we identify the regimes that control the melting dynamics, and provide a physical interpretation based on the competition between advection by the external flow and diffusive transport within the boundary layer.

The paper is organised as follows. Section~\ref{sec:setup}  introduces the numerical setup, governing equations and main control parameters of the problem. Section~\ref{sec:results} discusses the melting dynamics, distinguishing between slow (\S\ref{sec:slow_osc}) and fast   (\S\ref{sec:fast_osc}) oscillation regimes, and the role of the Stefan number (\S\ref{sec:effect_Stefan}). Finally, in section~\ref{sec:conclusions} we draw the conclusions and provide perspectives.

\begin{figure}
\centering
\includegraphics[width=1\textwidth]{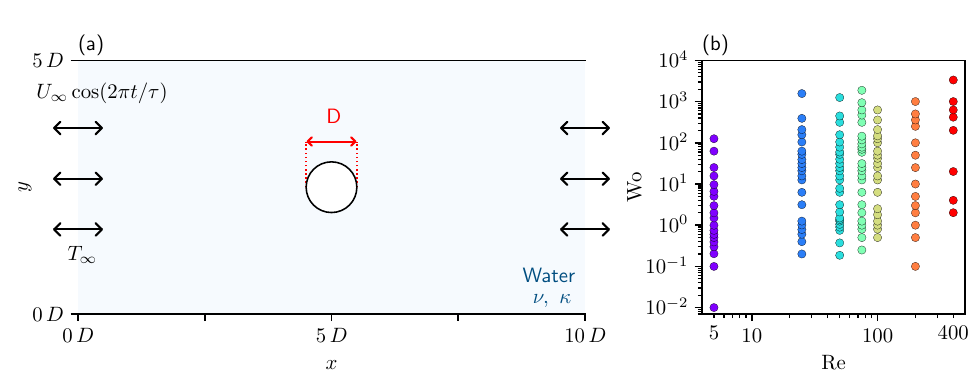}
\hfill
  \caption{(a) Set-up of the simulation. A fixed ice disk of size $D$ is placed at the centre of the domain, immersed in fresh, warm water. The channel is open on both ends, where both an inflow\textendash{}outflow condition, and far-field temperature are imposed. (b) Region of the Womersley number\textendash{}Reynolds number parameter space explored.}
\label{fig:setup}
\end{figure}

\section{Numerical setup}
\label{sec:setup}
Using two dimensional (2D) direct numerical simulations, we study the melting dynamics of an ice disk of initial diameter $D$ placed at the centre of a channel of height $5D$ and horizontal span $10D$. The ice is initially at its melting temperature $T_m$, and the channel is filled with fresh water at a temperature $T_\infty$. 
On both ends of the channel the temperature is set as uniform $T_\infty$, and the inflow and outflow velocity are set as $U_\text{in} = U_\infty \cos(2\pi t/\tau)$ by means of a penalty force, where $\tau$ is the oscillation period.
To focus only on the effect of forced convection, we neglect all buoyancy effects. The phase change is modelled by means of the phase-field method, using the same methodology as in references \citep{Yang2023,Yang2024,Yang2024_JFM,Favier2019, Hester2020}, where a scalar field $\phi$ transitions smoothly from $0$ in the liquid to $1$ in the solid, which evolves in time and space. The position of the ice-liquid interface is defined implicitly as $\phi = 0.5$.
We solve the Navier--Stokes and the continuity equations for the incompressible velocity field, the advection-diffusion equation for the temperature, and the evolution equation for $\phi$. In non-dimensional form these equations read

\begin{align} 
    \frac{\partial}{\partial t} \bm{u} + (\bm{u} \cdot \bm{\nabla}) \bm{u} &= -\bm{\nabla} p + \frac{1}{\text{Re}}\Big (\bm{\nabla}^2 \bm{u} - \frac{\phi \bm{u}}{\eta}\Big),\label{eq:momentum}\\
    \frac{\partial}{\partial t} \theta + (\bm{u} \cdot \bm{\nabla}) \theta &=  \frac{1}{\text{Re Pr}}\bm{\nabla}^2 \theta + \frac{1}{\text{Ste}}~ \frac{\partial}{\partial t} \phi,\label{eq:temperature}\\
    \frac{\partial}{\partial t} \phi &= \frac{6}{5} \, \frac{\text{Ste}}{ C\,  \text{Re Pr}} \Big[\bm{\nabla}^2 \phi - \frac{1}{\varepsilon^2} \phi(1 - \phi)(1 - 2 \phi + C\theta)\Big],\label{eq:phasefield}
\end{align}
where $\bm{u}$ is the incompressible (i.e. $\bm{\nabla} \cdot \bm{u} = 0)$ velocity field, in units of the inflow amplitude $U_\infty$, and $\theta$ is the non-dimensional temperature difference, defined as 
\begin{equation}
\theta = \frac{T-T_m}{T_\infty-T_m} = \frac{T-T_m}{\Delta T}.
\end{equation}
Lastly, $p$ is the pressure in units of $\rho_0\,U_\infty^2$, with $\rho_0$ the density of the fluid. The control parameters of the problem are the Reynolds number $\text{Re}$, the Prandtl number $\text{Pr}$, and the Stefan number $\text{Ste}$, defined as
\begin{equation}
    \text{Re} = \frac{U_\infty D}{\nu}, \quad \text{Pr} = \frac{\nu}{\kappa}, \quad \text{Ste} = \frac{c_p \Delta T}{\mathcal{L}}.
\end{equation}
Here $\nu$ and $\kappa$ are the kinematic viscosity and thermal diffusivity, respectively, and assumed to be equal in the solid and liquid phases, $c_p$ is the specific heat capacity, and $\mathcal{L}$ is the latent heat of melting.
The additional control parameter of the problem corresponds to the Womersley number $\text{Wo}$, 
\begin{equation}
    \text{Wo} = \frac{2\pi}{\tau} \frac{D^2}{\nu},
\end{equation}
representing the non-dimensional period of the oscillating inflow $\tau$, thus high values of $\text{Wo}$ correspond to a faster oscillation.

In eq.~\eqref{eq:phasefield} $\varepsilon$ represents the diffuse interface thickness (set to the mean grid spacing), $\eta$ is the non-dimensional damping time, the time over which the velocity tends to the velocity in the solid (set following \citet{Hester2020}). The coefficient $C$ is the phase mobility parameter, arising from the Gibbs--Thomson relation and here fixed at $C=10$. This parameter sets the variation of the melting temperature as a function of the local curvature of the interface. For a disk whose surface curvature is initially constant, this effect is negligible, and the chosen value reflects it.
Equations \eqref{eq:momentum}--\eqref{eq:phasefield} are solved using the second-order, staggered, finite-difference solver AFiD \citep{Verzicco1996, Ostilla2015, VanderPoel2015}. The top and bottom walls of the domain are adiabatic, and free-slip boundary conditions are imposed. All of the fields are solved in a uniform mesh in both directions using $N$ grid points in the vertical direction and $2N$ in the horizontal. The field $\phi$ is also solved in a uniform mesh using $N_r$ points in the vertical direction (and accordingly, $2N_r$ in the horizontal). We performed a grid convergence study for the case $\text{Re}=100$ and found that $N=384$ is sufficient to accurately solve the melting dynamics (consistent with the grid convergence reported in \citet{Yang2024_JFM}, using the same code as the one used in this study). For a different value of the Reynolds number $\text{Re}_\text{new}$, $N_\text{new}$ was chosen such that $N_\text{new} = N_\text{old} \sqrt{\text{Re}_\text{new}/\text{Re}_\text{old}}$, with $\text{Re}_\text{old}=100$. The number of grid points in the refined mesh is increased from $N_r=4N$ up to $N_r=12N$ as the Womersley number increases.

To reduce the number of free parameters, throughout this study $\text{Pr}$ is fixed and equal to $7$, and $\text{Ste} = 0.25$, values which correspond to fresh water at $20\degreecelsius$. We discuss the effect of varying the Stefan number in section~\ref{sec:effect_Stefan}.

\section{Results}
\label{sec:results}

\begin{figure}
\centering
\includegraphics[width=1\textwidth]{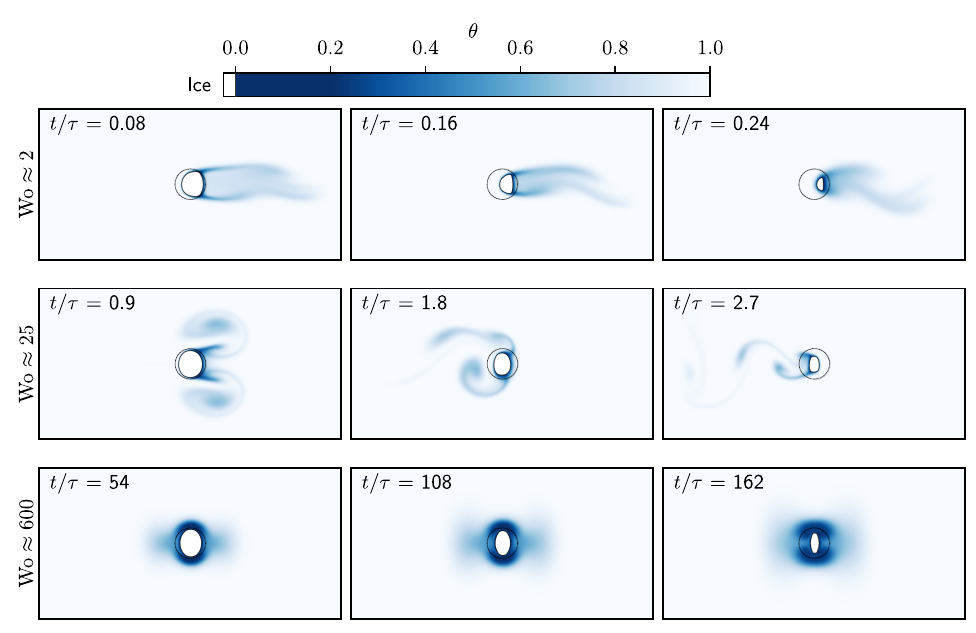}
\hfill
  \caption{
  Evolution of the non-dimensional temperature field $\theta$ (in colour) for $\text{Re} = 100$ in the full numerical domain. Each row represents a different value of $\text{Wo}$, and the time is measured in units of the external flow period $\tau$, which allows direct comparison of the flow across different cases, and highlights whether flow reversal has occurred during the melting process.
  The initial circular contour of the disk is indicated for reference with dashed lines.
  At $t/\tau=0$ the flow is from left to right.
  }
\label{fig:Re100_snapshots}
\end{figure}

Figure \ref{fig:Re100_snapshots} shows the evolution of the system for $\text{Re} = 100$, for three different values of $\text{Wo}$ (see also the supplementary videos). 
The top row shows snapshots of the temperature field and the ice shape for $\text{Wo} \approx 2$, where the oscillation is slow. The disk has melted almost completely before the flow has reversed direction (see the right-most panel, at $t=0.24\tau$), and the morphology of the ice looks similar to that of an object melting in a steady, unidirectional flow \citep{Yang2024_JFM}. The ice presents a front\textendash{}back asymmetry, with a rounded shape on the front-facing face, and a flatter back, consistent with results reported in the literature for similar systems affected by a steady flow \citep{Ristroph2012, Moore2013, Rycroft2016, Moore2017, Hewett2019}. The centroid of the ice shifts towards the right, indicating that melting is faster on the side facing the incoming flow (from left to right up to $t/\tau=0.25$).
In the middle row, corresponding to $\text{Wo} \approx 25$, as the object melts, and the flow reverses direction, the meltwater folds back onto the object. The shape remains more symmetric as compared to the slowly oscillating case.
The bottom row corresponds to a fast oscillating flow, for $\text{Wo} \approx 600$. In this case, the meltwater remains close to the object's surface throughout its entire evolution, shielding it from the warmer ambient water, and delaying its melting.
We observe a qualitatively similar behaviour for both smaller, and larger values of the Reynolds number (as it can be seen in the videos provided as supplementary material): a melting dynamics that resembles that in a steady flow for slow oscillations, cold meltwater folding back onto the ice at intermediate $\text{Wo}$, and shielding by meltwater at high oscillations. Although the structure of the wake and how it destabilises depends on $\text{Re}$, as it is known for flow past bodies \citep{Tritton2012}, this seems to play a secondary role in determining the overall melting dynamics in the slow and fast oscillation regimes.

\begin{figure}
\centering
\includegraphics[width=1\textwidth]{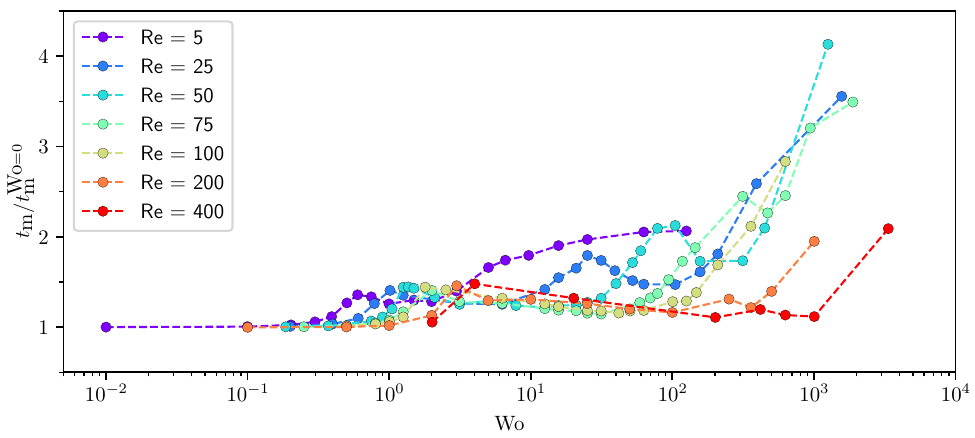}
\hfill
  \caption{Melting time, in units of the melting time in a steady flow, as a function of the Womersley number, for the different Reynolds considered, indicated by the colours.}
\label{fig:tmelt_vs_Wo}
\end{figure}

In figure~\ref{fig:tmelt_vs_Wo} we compare the melting time of the object $t_\text{m}$ (the time it takes for it to melt completely) with the melting time for the $\text{Wo} = 0$ case $t_\text{m}^{\text{Wo}=0}$, i.e., the melting time in a steady, unidirectional flow.
For low values of $\text{Wo}$ the oscillating flow is slow enough that the ratio of melting times approaches $1$. This is to be expected, as at the beginning of the evolution the external flow starts as $U_\text{in} = U_\infty$. Then, for a small frequency, the external flow which be approximately constant and with an amplitude close to $U_\infty$ throughout the disk melting.
As the frequency of the external flow increases (larger $\text{Wo}$), $t_\text{m}$ becomes larger than in the steady case, with a local peak occurring systematically for all explored values of $\text{Re}$. The value of $\text{Wo}$ at which this peak occurs is also $\text{Re}$-dependent. As $\text{Wo}$ increases further, a slight decrease of the melting time is observed, but $t_\text{m}/t_\text{m}^{\text{Wo}=0}$ remains greater than 1. For larger values of $\text{Wo}$ a sharp increase in $t_\text{m}$ is observed, systematically for all Reynolds number, and consistent with what is shown in the bottom row of figure~\ref{fig:Re100_snapshots}. Again, the value of the Womersley number where this transition occurs  is $\text{Re}$-dependent.

\subsection{Slow oscillations}
\label{sec:slow_osc}
\begin{figure}
\centering
\includegraphics[width=1\textwidth]{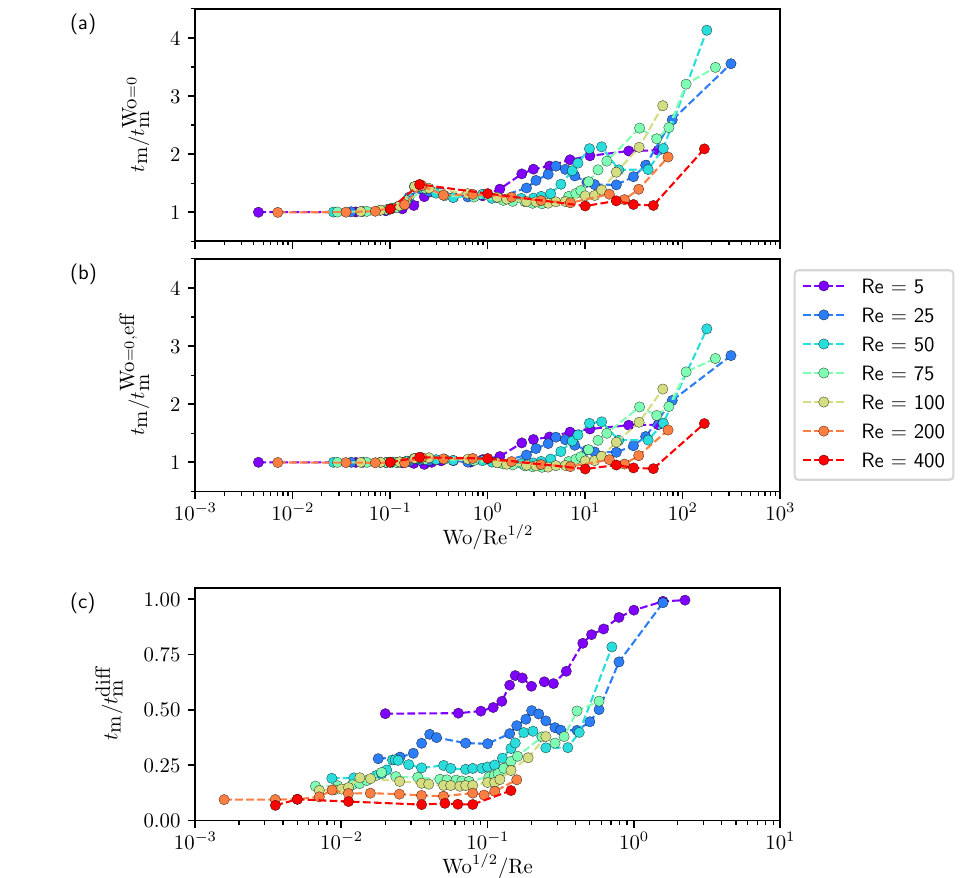}
\hfill
  \caption{
  (a) {\it Slow oscillations} \textendash{} Melting time in units of the melting time in a steady flow as a function of $\text{Wo}/\text{Re}^{1/2}$. The horizontal axis is shared with panel (b), which shows $t_\text{m}$ compensated by the melting time in an effective steady flow.
  (c) {\it Fast oscillations} \textendash{} Melting time in units of melting time by pure diffusion, as function of $\text{Wo}^{1/2}/\text{Re}$.
  The legend for all panels is indicated on the right.
  }
\label{fig:low_high_Wo}
\end{figure}

In the slow oscillation regime, the peak at low Womersley reported in figure~\ref{fig:tmelt_vs_Wo} can be collapsed across the different $\text{Re}$ by compensating the horizontal axis by $\text{Re}^{1/2}$, as is shown in panel (a) of figure~\ref{fig:low_high_Wo}. In this way, the peak occurs at approximately $\text{Wo}/\text{Re}^ {1/2} \approx 0.2$. 
As the external flow has the functional form of a cosine, this corresponds to the time interval where the flow has not reversed direction, i.e. when $t_\text{m} < \tau/4$.
The average Nusselt number $\overline{\text{Nu}}$, the surface-averaged heat flow, is related to the melting time $t_\text{m}$ via the relation
\begin{equation}
    \overline{\text{Nu}} \propto \frac{D^2/\kappa}{t_\text{m}} \frac{1}{\text{Ste}}.
    \label{eq:Nu_definition}
\end{equation}
In forced convection within a steady flow $\overline{\text{Nu}}$ has been reported to be proportional to $\text{Re}^{1/2} \,\text{Pr}^{1/3}$, which holds also for melting of ice
\citep{Yang2024_JFM,Xue2026}. Combining this with eq.~\eqref{eq:Nu_definition}, one obtains
\begin{equation}
    \frac{t_\text{m}^{\text{Wo}=0}}{\tau} \propto \frac{\text{Wo}}{\text{Re}^ {1/2}} \frac{\text{Pr}^ {2/3}}{2\pi\, \text{Ste}}.
\end{equation}
In our simulations shown in figures~\ref{fig:tmelt_vs_Wo} and~\ref{fig:low_high_Wo}, the Prandtl and Stefan numbers are constant and equal to $\text{Pr} = 7$ and $\text{Ste}=0.25$.
Then, the ratio $\text{Wo}/\text{Re}^{1/2}$ can be interpreted as the ratio between the inflow period of oscillation, and $t_\text{m}^{\text{Wo}=0}$. The peak at $\text{Wo}/\text{Re}^{1/2} \approx 0.2$ corresponds to $t_\text{m}^{\text{Wo}=0} \approx 0.2\, \tau$, i.e. the melting time of an ice disk in a steady flow is shorter than the time the external flow reverses direction. The proportionality factor in this scaling is not expected to be strictly of order 1, and indeed empirical correlations for the Nusselt number typically include prefactor that depend on the specific geometry and flow configuration. Nevertheless, the reported collapse of the data indicates that $\text{Wo}/\text{Re}^{1/2}$ gives the correct scaling for the transition, even if pre-factors are not precisely captured.

In the cases where the object has melted completely before the external flow has reversed direction, the ice disk is being subjected to a uni-directional flow whose amplitude is decreasing in time. It is natural to ask whether this situation can be described by an equivalent steady flow. For that, we define an effective Reynolds number by taking an average velocity during the melting process as follows
\begin{equation}
    \text{Re}_\text{eff} = \frac{U_\infty D}{\nu} \, \frac{1}{t_\text{m}}\int\limits_0^{t_\text{m}} \left\lvert \cos \left(\frac{2\pi\,t}{\tau}\right) \right\lvert\, dt.
\end{equation}
Following the same arguments as before, the melting time in a steady flow with Reynolds number $\text{Re}_\text{eff}$, $t_\text{m}^{\text{Wo}=0,\,\text{eff}}$, is estimated as
\begin{equation}
    \frac{t_\text{m}^{\text{Wo}=0,\,\text{eff}}}{\text{Re}_\text{eff}^{1/2}} \approx \frac{t_\text{m}^{\text{Wo=0}}}{\text{Re}^{1/2}}.
\end{equation}
The inset (b) of figure\,\ref{fig:low_high_Wo} shows $t_\text{m}$ now normalised by $t_\text{m}^{\text{Wo}=0,\,\text{eff}}$. By doing this, the peak at $\text{Wo}/\text{Re}^ {1/2} \approx 0.2$ disappears, and the ratio $t_\text{m}/t_\text{m}^{\text{Wo}=0,\,\text{eff}}$ is approximately $1$. Notably, this occurs up to $\text{Wo}/\text{Re}^ {1/2} \approx 1$, which corresponds to $\tau \approx t_\text{m}^{\text{Wo}=0}$. This indicates that the melting in the presence of an oscillatory flow behaves like that in a steady flow, provided an effective Reynolds number is considered, allowing to characterise the melting time within the framework of forced convection up to $\text{Wo}/\text{Re}^{1/2} \approx 1$.

\subsection{Fast oscillations}
\label{sec:fast_osc}
As observed qualitatively in the third row of figure~\ref{fig:Re100_snapshots}, when the frequency of oscillation of the external flow increases, the cold meltwater remains in the surroundings of the object, only moving far from the object at very long timescales, which results in an increase of the melting time of the ice disk.
Recalling that in the current study buoyancy effects have been neglected, it is then appropriate to compare $t_\text{m}$ with the melting time in a purely diffusive regime $t_\text{m}^ \text{diff}$, in the absence of a flow where melting occurs simply due to heat diffusion.
This is shown in figure\,\ref{fig:low_high_Wo}(c), as a function of $\text{Wo}^{1/2}/\text{Re}$. 
We observe that when $\text{Wo}^{1/2}/\text{Re} \approx 10^{-1}$, the melting time for all the different $\text{Re}$ considered increases, to reach the melting-by-diffusion limit when $\text{Wo}^{1/2}/\text{Re} \approx 1$, this being being particularly noticeable for the smallest Reynolds numbers considered.
Note that exploring the regime towards higher values of $\text{Wo}^{1/2}/\text{Re}$ is computationally very demanding, and the current maximum values are at the limits of the computational demands accessible to us.

To understand this transition, we first recall that in an oscillating flow with angular frequency $\omega = 2\pi/\tau$, a momentum oscillating (or Stokes) boundary layer develops, with typical thickness $\delta_S = \sqrt{2\nu/\omega}$ \citep{Batchelor2000}. Secondly, the tidal displacement $\Delta$ in our problem, i.e. the typical distance travelled by a fluid parcel in a period of oscillation, is proportional to $U_\infty\tau$. The ratio between the two can be re-written in terms of input parameters of the problem,
\begin{equation}
    \frac{\delta_S}{\Delta} = \frac{\sqrt{2\nu}}{\sqrt{\omega}\, U_\infty \tau} = \sqrt{\frac{2}{2\pi}} \frac{\sqrt{\nu}}{U_\infty\sqrt{\tau}} = \frac{\sqrt{2}}{2\pi} \frac{\text{Wo}^{1/2}}{\text{Re}}.
\end{equation}
When $\Delta < \delta_S$ (i.e., $\text{Wo}^{1/2}/\text{Re} > 1$), the advection by the external flow is not strong enough to remove cold water from within the thermal boundary layer, nor to bring in warm, ambient water during one period of oscillation. The melt water will remain close to the ice object, thermally shielding it and delaying the melt. Indeed, under these conditions, diffusion has sufficient time to act and homogenise the temperature within the boundary layer. As the temperature difference between the surface and the surrounding fluid decreases, the heat flux decreases and the melting time is extended.

\subsection{Effect of varying the Stefan number}
\label{sec:effect_Stefan}
The Stefan number, which compares the sensible heat to the latent heat of melting, is a measure of how fast or slow the melting process is compared to heating the ambient. For a fixed temperature difference between the ice and the ambient $\Delta T$, the smaller $\text{Ste}$, the larger the latent heat of fusion, so more heat is needed to be supplied by the ambient to the ice for it to change phase. Thus, the Stefan number naturally introduces an additional time scale into the problem. In this section we explore how this is coupled to the external flow oscillation period $\tau$. We consider two Reynolds numbers, $\text{Re} = 50$ and $ \text{Re} = 200$, for different values of $\text{Wo}$, including the steady flow case, and vary $\text{Ste}$ in the range $[1/40, 1]$.

\begin{figure}
\centering
\includegraphics[width=0.99\textwidth]{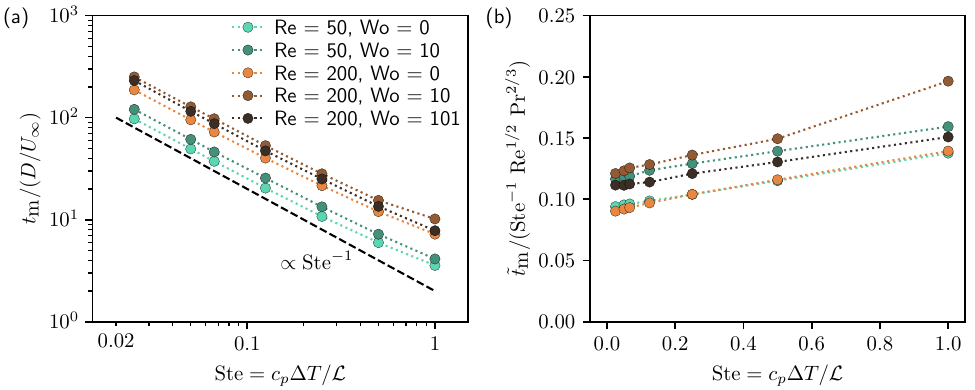}
\hfill
  \caption{Melting time dependence on the Stefan number. (a) Normalised melting time $\tilde{t}_\text{m}\equiv t_\text{m}/(D/U_\infty)$ as a function of $\text{Ste}$, for two different Reynolds number, for different values of $\text{Wo}$, and for the steady flow case ($\text{Wo}=0$). The dashed line indicates a power law with exponent $-1$ for reference. (b) Normalised melting time compensated by the scaling prediction stemming from forced convection. Labels are the same as in panel (a).}
\label{fig:Stefan_dependence}
\end{figure}

In figure~\ref{fig:Stefan_dependence}(a) we show the non-dimensional melting time $\tilde{t}_m \equiv t_m/(D/U_\infty)$ as a function of the Stefan number for the different $\text{Re}$ and $\text{Wo}$ explored. The data seem to follow a power-law decay with an exponent close to $-1$, shown as reference in the figure with a dashed line. Following equation~\ref{eq:Nu_definition} and with the same scaling laws as before, it follows that
\begin{equation}
    \tilde{t}_m \propto \frac{1}{\text{Ste}}\, \text{Re}^{1/2}\, \text{Pr}^{2/3},
\end{equation}
which have been shown to describe well the Stefan number dependence of melting in forced convection \citep{Yang2024_JFM, Xue2026}. In the case of melting due to an oscillating flow the same scaling behaviour seems to hold, regardless of the value of $\text{Wo}$. When plotting the compensated melting time in figure~\ref{fig:Stefan_dependence}(b), we observe that indeed the scaling describes with reasonable accuracy our data, with a weak $\text{Ste}$ dependence, and with some spread in $\text{Wo}$.
For larger values of $\text{Wo}$ than the ones shown here, when melting is dominated by diffusion, finite-domain size effects become increasingly important as the Stefan number decreases. 
In this regime, the melting process is sufficiently slow that the cold meltwater diffuses across the entire domain before the ice has fully melted. As a result, the temperature field becomes influenced by the finite-size of the domain, leading to an additional reduction in heat transfer, and a further increase in melting time.

\section{Conclusions}
\label{sec:conclusions}

We have investigated the melting dynamics of an ice disk subjected to an external oscillatory flow using two-dimensional direct numerical simulations, neglecting buoyancy effects. By varying the Reynolds number $\text{Re}$ and the Womersley number $\text{Wo}$, we have identified two distinct regimes of melting, controlled by the competition between the advection from the external flow, and diffusion within the boundary layer.

For slow oscillations, $\text{Wo}/\text{Re}^{1/2} \le \mathcal{O}(1)$, the melting is well described by an effective steady flow. In this limit, the ice disk melts within a fraction of the oscillation period, before the flow has reversed direction. The melting time can be described using classical forced convection scaling laws, and introducing an effective Reynolds number based on the average velocity over the melting period shows that the oscillation does not change the fundamentals of heat transfer in this regime, allowing us to fully predict the melting time in this regime.

For fast oscillations the melting dynamics changes significantly with respect to the steady-flow case, displaying a sharp increases in melting time reaching the purely diffusive limit when $\text{Wo}^{1/2}/\text{Re} \ge \mathcal{O}(1)$. When the displacement of a fluid parcel over one oscillation period $\Delta = \mathcal{O}\left( U_\infty \tau\right)$ is comparable to the oscillatory boundary layer $\delta_s\approx \sqrt{2\nu/\omega}$, the oscillation is not strong enough to renew the fluid within the thermal boundary layer, which extends the melting time.

Finally, we have shown that the dependence of the melting time with the Stefan number follows the same scaling as in steady forced convection, indicating that the role of latent heat remains mostly decoupled to the oscillatory dynamics.

Our results show that oscillations can inhibit heat transfer when failing to renew the fluid within the boundary layer, which is also a consequence of the absence of buoyancy effects. In fact, in the presence of natural convection, density differences between the cold meltwater and the warmer ambient fluid would induce buoyancy-driven flows, which would enhance the removal of meltwater from the vicinity of the ice surface, which would modify the fast oscillation regime described here. Investigating the competition between unsteady, oscillating forced convection and natural convection represents an interesting direction for future work.

\section*{Acknowledgments}
We thank D. Lohse for fruitful discussions.
This work has been  funded by the European Union (ERC, MeltDyn, No. 101040254).
Numerical resources were provided by the EuroHPC Joint Undertaking for awarding the project EHPC-REG-2023R03-178 to access the EuroHPC supercomputer Discoverer, hosted by Sofia Tech Park (Bulgaria). We also acknowledge the Dutch national e-infrastructure with the support of SURF Cooperative.


\bibliography{ms}

\end{document}